\magnification=1200

\centerline{\bf FACT \& FANCY IN NEUTRINO PHYSICS II}
\vskip 0.5 truecm
\centerline{ Sheldon Lee Glashow}
\centerline{\it Physics Department, Boston University}
\centerline{\it 590 Commonwealth Avenue, Boston, MA 02215}
\centerline{Email: slg@bu.edu}

\vskip 1 truecm
{\medskip\narrower\narrower\centerline {ABSTRACT } 
\noindent
This brief and
opinionated essay evolved from my closing talk at the Tenth International
Workshop on Neutrino Telescopes, held in Venice in March 2003. Portions were
inspired by several excellent
 presentations at the Workshop. Other scattered comments
about neutrino physics relate to variations of the seesaw model yielding the
FGY ansatz, or to those yielding significant suppressive mixing
of neutrino amplitudes.\bigskip}

I am honored to have been chosen to give the closing address at this
Workshop.  The late and beloved Viki Weisskopf described the privilege of
being a physicist. Milla Baldo-Ceolin, on ten occasions, has given us the
privilege of practicing our art in La Serendissima. Let me begin by thanking
Milla and her staff for making these wondrous Venetian workshops
possible.\medskip

The original F\&FiNP was presented as a Harvard Colloquium in the form of a
play in December 1973, just after neutral currents were found and just
before  the dramatic discovery of the curiously called $J/\Psi$
particle. Our  play was later published in the Reviews of Modern Physics [1].
The  cast consisted of:

{\smallskip \obeylines
Alvaro De R\'ujula: \ Moderator,  an Experimental Physicist
Howard Georgi: \  Computer, one that can talk
Helen R.  Quinn:\  Speaker, a Conservative Theorist
and me: \ Model-Builder, a not-so-conservative Theorist\smallskip}

\noindent The  plot centered upon the  
exciting new data then emerging on deep-inelastic
lepton scattering, and their interpretation in terms of a naive quark model,
but one involving quarks yet undiscovered: those with {\it charm\/} (which do
exist) and those with {\it fancy\/} (which do not).  It was a heady time in
the history of particle physics, somewhat confused by Rubbia's soon-to-vanish
`high-y anomaly.'  Milla's request
 for a reprise of Fact and
Fancy is impossible to fulfil in these more tepid  days,
 but as I attempt to recall its spirit
 please remember that Facts
refer to  suppositions that are true, Fancy to those that rest on no solid
ground.  \medskip

 Colleagues occasionally ask why I never claimed credit for the invention of
the seesaw model of neutrino masses --- that is, the scheme by which neutrino
masses arise from an interplay beteen Higgs-induced Dirac masses involving
three weak doublet neutrinos and three singlet states, and large bare
Majorana masses of the singlets. In lieu of staking a claim, let me offer a
chronological list of the earliest published discussions of the seesaw model:

{\smallskip\obeylines

1)  Tsutomu  Yanagida in {\it Proc. Workshop on Unified Theories \&c.,}
\indent [Feb. 13-14, 1979], eds. O. Sawada and A. Sugamoto (Tsukuba,1979) p.95.
\smallskip

2) S.L. Glashow, in {\it Quarks and Leptons, Carg\`ese} [July 9-29,1979], 
eds. M. L\'evy, {\it et al.,} (Plenum, 1980, New York), p. 707.

\smallskip

3) M. Gell-Mann, P. Ramond, and R. Slansky, in {\it Supergravity,} [Sept. 27-29
1979],  eds. D. Freedman {\it et al.,} (North Holland, 1980, Amsterdam).
\smallskip

4) R.N. Mohapatra and G. Senjanovi\'c, Phys. Rev. Lett. 44 (1980) 912.
\medskip}
 
\noindent In my 1979 Carg\`ese talks, I wrote: ``Consider the effect of
[neutrino] mixing on the distribution of neutrinos produced by cosmic rays.
Upward directed neutrinos have a trajectory of $\sim 10^{4}\;$km while
downward directed neutrinos travel only $\sim 10\;$km... It is possible that
the next world-shaking developments in particle physics will emerge from such
experiments.''  Little did I realize that I would  wait almost two
decades before the anticipated atmospheric neutrino oscillations would be
detected. If only Bruno Pontecorvo could had seen how far we
have come toward understanding the pattern of neutrino masses and mixings!
Way back in 1963 he was among the first have envisaged the possibility of
neutrino flavor oscillations.  For that reason, the analog to the
Cabbibbo-Kobayashi-Maskawa matrix pertinent to neutrino oscillations should
be known as the PMNS matrix, to honor four neutrino visionaries: Pontecorvo,
Maki, Nakagawa, and Sakata.\medskip

{\bf A plea!}  The mixing angles appearing in 
standard parametrizations of both the PMNS matrix and the CKM matrix
 are
usually designated  by $\theta_{12}$ (solar/Cabibbo), 
$\theta_{23}$ (atmospheric/$b\rightarrow c$) and
$\theta_{13}$ (subdominant/$b\rightarrow u$). 
It is awkward and absurd  to use two indices where one
would do. Therefore, I prefer, recommend and
shall hereafter use a  simpler and more compact  notation:
$$\theta_1\equiv \theta_{23},\qquad \theta_2\equiv\theta_{13}, \qquad
\theta_3\equiv \theta_{12}.$$
What we have managed thusfar to learn about these parameters 
(and the CP violating phases $\delta$) is rather roughly summarized 
 in the following table: 
\def\h{\hbox}
$$\hbox{\vbox{
\hbox{\bf Parameter}
\hbox{}
\hbox{$\sin{\theta_1}$}
\hbox{$\sin{\theta_2}$}
\hbox{$\sin{\theta_3}$}
\hbox{$\delta$} }\qquad\vbox{
\hbox{\bf Quarks}\h{}\h{0.04}\h{0.004}\h{0.22}\h{$\sim 1$} }
\qquad\vbox{\h{\bf Leptons}\h{}
\h{$\sim \sqrt{2}/2$}\h{$\le 0.16$}\h{$\sim 0.55$}\h{??}}}$$

{\bf A question! } How much better must 
we strive to determine  these parameters,
about which our theories are so sadly reticent? 
 For the quark sector, the answer primarily involves the two unitarity
relations:
$$\eqalign{
V_{ud}\,V_{ub}^* + V_{cd}\,V_{cb}^* + V_{td}\,V_{tb}^* =& \;
 0,\quad {\rm and}\cr
|V_{ud}|^2 + |V_{us}|^2 + |V_{ub}|^2 \equiv 1-\Delta=& \; 1.\cr}$$
The first of these 
  is usually called  the `unitarity triangle.' Current
measurements indicate  that the triangle  inequalities are obeyed with
 the angle between the latter two legs
  given by  $\sin{2\beta}=0.78\pm 0.08$. Further data is required
to confirm the view 
that standard-model CP violation, {\it i.e.,} that implied 
solely by the
complexity of the CKM matrix,  offers a correct and complete description of
all observable 
CP-violating phenomena in both the kaon and $B$-meson sectors.
(Here we ignore the so-called strong CP problem.)
\medskip   

The second relation poses a small puzzle.
Wilkinson [2], from a recent
simultaneous  analysis of several super-allowed Fermi transitions, obtains
$$\Delta= 0.0004\pm 0.0017\,,$$
which is in excellent agreement with CKM universality, 
whereas  Abele [3], from a new
measurement of the neutron lifetime, obtains
$$\Delta = 0.0083\pm 0.0028,$$
which is  a 3-sigma discrepency from theory.
At least one of these two estimates must be flawed. Furthermore, I have
heard  that the long-accepted value $|V_{us}|\simeq 0.22$ may be
challenged by high-statistics studies of kaon decay.
 The current situation is 
 confused. However, a confirmed  departure
from CKM universality (whether positive or negative),
should there be one,  would be  decisive evidence
for  physics beyond the standard model.   \medskip

On that engaging note, let us look  to the leptons.  First off, we had best
verify our three-state description of neutrino oscillations and
convince ourselves that the ugly construct of `sterile neutrinos' can be
safely abandoned. This done, with what precision must the leptonic mixing
angles be determined? I would argue as follows. Being aware of no compelling
theoretical reason for any of the leptonic angles $\theta_i$ to assume
special values, such as zero or $\pi/4$, I would be satisfied if atmospheric
neutrino oscillations (which are known to be nearly maximal) could be
measured sufficiently well to bound $\theta_1$ away from $\pi/4$ with
reasonable certainty. Similarly, solar oscillations (which are strongly
favored to be less than maximal) should be measured sufficiently to bound
$\theta_3$ away from $\pi/4$ with 5-sigma certainty. Finally, I would be
satisfied if the subdominant angle $\theta_2$ could be bounded away from zero
with 5-sigma certainty. Once these benchmarks are met, I would argue that
further measurements of neutrino oscillation phenomena would be pointless,
until and unless theorists can provide further guidance.\medskip
    
That leaves the question of $\delta$, the parameter governing CP violation in
neutrino physics. Discussions at this Conference suggest  that the
cost of bounding $\delta$ from zero would be enormous. It has been said that
a reliable measurement of $\delta$  would require the
construction of a multi-billion euro
muon factory, for which this  task would be the sole
{\it raison d'\^ etre.} 
 I believe that such a  facility
is not affordable by our  presently impoverished discipline.
Unless a cheaper route to $\delta$ can be found,
there are  too many less costly but equally important challenges remaining
in neutrino physics that should command our limited funds.
Among them are the following: \smallskip

\item{$\bullet$} Pinning down the leptonic mixing angles, as described above. 

\item{$\bullet$} Searching for  neutrinoless double beta decay.

\item{$\bullet$} Studying the tritium endpoint so as to constrain 
neutrino masses.

\item{$\bullet$} Measuring the two squared-mass differences among neutrinos.

\item{$\bullet$} Distinguishing the normal from  
the inverted neutrino mass spectra.

\item{$\bullet$} Resolving the LSND anomaly and confirming the 3 active
neutrino scenario.

\item{$\bullet$} Testing CPT for neutrinos, {\it e.g.,} by comparing solar and
Kamland data.

\item{$\bullet$}  Improving  the already astonishing cosmological limit
 on the sum of the neutrino masses, such  as 
was eloquently described by Prof. Pastor at this meeting.\bigskip

 {\bf  Seesaws! } 
Let me return to the wondrous mechanism which is somewhat  the
theme of my  talk. It is a neat way
to generate neutrino masses with a minimum of new architecture.
At the same time,  the seesaw can address the
mystery  of universal baryon asymmetry. 
Professor Buchm\"uller, at this meeting, stressed the virtue of the
seesaw 
to  implement leptogenesis and consequent baryogenesis. Indeed,
he claimed  that such a scheme yields  a powerful
constraint on the sum of the
squares of the light neutrino masses. He works in the context of three heavy
singlet neutrino states, where  one of them
is significantly lighter than the
others.  Otherwise, the model in an unconstrained seesaw with Dirac masses
arising from a single Higgs boson. It is interesting to note that
leptogenesis, in this model, occurs  entirely independently
 of the parameters of the
PMNS matrix. In particular, the requisite CP violation is controlled by
otherwise inaccessible `Majorana phases,' and not by the $\delta$ parameter.  
(Early references to these phases, and indications of their unobservability,
 may be found in ref.  [4].)\medskip

 The situation is more constrained  and perhaps more interesting
should  we adopt the simple Frampton-Glashow-Yanagida 
FGY  {\it ansatz\/} [5] in which there are 
just two heavy singlets ($N_i$),  and  the 
neutrino mass terms (in an abbreviated
notation)  
take the   special form:
$$ (a\nu_e + b\nu_\mu)\langle h\rangle\,N_2 +
(c\nu_\mu + d\nu_\tau)\langle h\rangle\,N_1 
+M_1\,N_1N_1 + M_2\,N_2N_2,$$
or a related form in which the roles of $\nu_\mu$ and $\nu_\tau$ are
reversed. The Higgs vev is denoted by $\langle h\rangle$. 
The Yukawa couplings 
 $a,b,c,d$ are arbitrary complex numbers, but we may choose
the flavor phases so as to make all but one of them real. Consequently,
and for either version of our  ansatz, exactly one
convention-independant phase  controls both 
leptogenesis (via $N$ decay) and CP violation in the 
neutrino sector.
Thus 
 the (unknown) sign characterizing CP violation in the
neutrino sector is correlated to  the (known) sign of the
baryon asymmetry of the universe. 
\medskip

Furthermore, and as noted by Raidal \&\ Strumia [6] (who have dubbed our
model
``the most minimal seesaw''),
the FGY  ansatz has four    potentially observable
consequences at low energy: 
\smallskip

\noindent (1)  One of the three neutrino masses must
vanish.
\smallskip
\noindent   (2) The inverted neutrino mass hierarchy is excluded. 
\smallskip

\noindent (3) The neutrino
flavor-mixing parameters are constrained by the relation:
$$ \sin{\theta_2}= \sin{2\theta_3}\big(\tan{\theta_1}\big)^{\pm 1}
\sqrt{{\Delta_s\over 4\Delta_a}}\,,$$
where the sign ambiguity (linked to the choice of ansatz) is not
phenomenologically significant because of the experimental result
$\tan{\theta_1}\approx 1$.  The parameters $\Delta_{a,s}$ are the squared
mass differences pertinent  to atmospheric and solar neutrino oscillations. 
Insofar as experiment strongly indicates  solar neutrino oscillations 
to be  far from maximal, this relation predicts  the subdominant angle
$\theta_2$ to be  large enough 
to make searches for CP violation in the neutrino
sector feasible.

\smallskip

\noindent  (4) The element of the neutrino mass matrix
responsible for neutrinoless double beta decay is given by
$$ M_{ee}= \sin^2{\theta_3}\sqrt{\Delta_s}\,,$$
which is small enough to pose a formidable challenge to experimenters who
would search for this rare or nonexistent process. The importance of this
search cannot be overstated. The detection of no-neutrino $\beta\beta$ decay
would prove that lepton number is not conserved, and  
would thereby exclude the
otherwise tolerable possibility (pointedly stressed by Jack Steinberger)
that neutrino masses are ludicrously hierarchic but purely Dirac. 
\medskip

The attentive reader will have noted
 that the elegant  and predictive
ansatz we  proposed is unnatural and unrenormalizable {\it per
 se.}  Without further ado, its form
 is not preserved by divergent radiative
 corrections. This difficulty can be remedied in several ways.  For example,
 Raby [7] claims to have  generated our  ansatz naturally in a
 supersymmetric context such as to preserve the correlation between the
 baryon asymmetry and observable CP violation in the neutrino sector. Another
 way to accomplish these goals  is sketched below.
\medskip

We assign a flavor quantum number $F$ to the light leptons as follows.  To
the (electron, muon, tau lepton) and its left-handed neutrino, we assign $F=
(1,0,-1)$, resp. We introduce three Higgs doublets, $h_i$ to account for
lepton masses. Their subscripts are identified with their flavor assignments:
 $F\equiv i= (0,\,-3,\,-4)$. In a similar fashion, 
the left-handed singlet neutrino $N_1$ is 
renamed $N_4$ and assigned $F=4$, while $N_2$ is 
renamed $N_3$ with 
$F=3$. It will be required that    
$F$ be  conserved by all dimension-4 terms in the
Lagrangian. However, $F$ is softly broken by the dimension-3
Majorana mass terms, 
 but in such a manner as to conserve $F$ modulo 2. Thus these terms are
constrained to be 
$M_3\,N_3^2 + M_4\,N_4^2$. The $M_i$ may be chosen to be real with no loss of
generality. I summarize 
below the other consequences of these flavor assignments:
\medskip

\noindent
$\bullet$ The charged lepton masses arise  from Yukawa couplings of
$h_0$ to light lepton states. They are necessarily
flavor diagonal in the basis we are
using. Note that $h_0$ has no $F$-conserving
 couplings involving  $N_i$, and that $h_i$ (for $i\ne 0$) do not contribute
to the charged lepton masses.  
\smallskip

\noindent$\bullet$ The allowed couplings of the light doublet states to the
$N_i$ are just those  required to reproduce the FGY ansatz:
$$ \big(\hat{a}\langle h_{-4}\rangle\, \nu_e + 
\hat{b}\langle h_{-3}\rangle\,\nu_\mu\big)N_3 +
\big(\hat{c}\langle h_{-4}\rangle\, \nu_\mu + \hat{d}
\langle h_{-3}\rangle\,\nu_\tau\big)N_4\,.$$ 
In this manner,  the structure   of the FGY  ansatz 
is preserved
by radiative corrections  (up to small finite terms) 
because it is protected by the softly-broken flavor symmetry, and  so also
are all of its observable consequences.

\bigskip

 {\bf Suppressive neutrino mixing!}
Two potentially  threatening  departures from\break
standard-model predictions regarding neutrino
physics
 are the three sigma NuTeV anomaly
[8] and the two sigma departure of the $Z^0$ invisible width from its expected
value [9].  In the former case, a ratio of neutral-current to charged current
cross sections is reported to be less than  its predicted value by $1.2\pm
0.4\,$\%.  In the latter case, the neutrino count is shy of three by
$0.016\pm 0.008$.  
\medskip

Many authors  have proposed, considered, criticized 
or rejected explanations of these
`anomalies'  involving  significant suppressive 
mixings between light and heavy neutrino
states [10]. Finding neither discrepancy convincing, we need make no comment
on this issue.  Nonetheless, several recent papers [11] address the ancillary
question of how this mixing can arise in seesaw models and what would be its
observable consequences.
The upshot 
of these analyses is that the effective flavor
eigenstates ({\it i.e.,} the doublet states coupled to $e,\,\mu$ and $\tau$,
 projected onto the space of light neutrino eigenstates) 
are neither normalized nor orthogonal. They satisfy the relation:
$$ \big(\nu_l^\dagger\cdot \nu_{l'}\big) = \delta_{ll'}
 - \Theta_{ll'}\,,$$
where $\Theta$ is a small non-negative
hermitean matrix. Ordinarily,  the
entries of  $\Theta$ are   given by 
ratios of mostly doublet  light neutrino masses to their mostly singlet
heavy counterparts. These  are typically less than $ 10^{-11}$ and 
therefore entirely negligible.   
This is the case, for example, for the FGA ansatz
discussed previously.

\medskip
It has been observed [11], however,  that 
carefully chosen values  of the seesaw parameters can lead to
very  much larger (and empirically relevant) $\Theta$'s.
For these pathological  cases,  the diagonal components of $\Theta$ can 
yield  significant suppressions of the various 
weak-interaction leptonic amplitudes involving neutrinos, 
such as might  ameliorate  the above-cited anomalies, or aggravate  the
previously-discussed disputed departure from
CKM universality. Similarly,
the  off-diagonal components of $\Theta$ 
can induce, via loop diagrams, 
otherwise forbidden processes such as $\mu\rightarrow e+\gamma$,
$\mu\rightarrow 3e$, and $\mu$--$e$ conversion.
For these contrived situations   
the entries of the matrix $\Theta$ are virtually independent of the
neutrino masses, although they are not entirely arbitrary. In any event, 
it would seem  worthwhile  to set 
direct experimental  upper bounds on these
possible departures from neutrino orthonormality and to delineate the
theoretical constraints upon them. 

\bigskip\bigskip
{\bf Acknowledgement: } I am grateful for the  enlightening comments of 
Professors A. Cohen, P. Frampton and P. Langacker. This  
 research was supported in part by the National Science
Foundation 
under grant number NSF-PHY-0099529.

\vfill\eject
%\vskip 1.5 truecm
%\bigskip\bigskip
{\bf References} 

\noindent[1] A. De R\'ujula {\it et al.,} Rev. Mod. Phys. 46(1974)391.

\noindent[2] D.H. Wilkinson, J. Phys. G; Nucl. Part. Phys. 29(2003)189.

\noindent
[3] Helmut Abele, in Rencontres de Moriond 2002, {\tt hep-ex/0208048.}

\noindent [4] A.  de Gouv\^ea, 
B. Kayser \& R.N. Mohapatra, {\tt hep-ph/0211394}.

\noindent
[5] P.H. Frampton, S.L. Glashow \& T. Yanagida, Phys. Lett. B548(2002)119.

\noindent
[6] M. Raidal \& A. Strumia, Phys. Lett. B553(2003)72.

\noindent [7] Stuart Raby, {\tt hep-ph/0302027.}

\noindent [8] K.S McFarland {\it et al.,} {\tt hep-ex/0205081};
 G.P. Zeller {\it et al.,} {\tt hep-ex/0207037.}

\noindent [9] The LEP Collaborations, the LEP
  Electroweak Working Group,\hfil\break
 and the  SLD Heavy Flavor Group, {\tt
    hep-ex/0212036.}

\noindent [10] {\it E.g., } 
 J. Bernard\'eu {\it et al.,}
  Phys. Lett B187(1987)303;\hfil\break
  A. de Gouv\^ea {\it et al.,} Nucl. Phys.
  B623(2002)395, {\tt hep-ph/0107156}; \hfil\break
K.S. Babu and J.C. Pati, {\tt
    hep-ph/0203029}; S. Davidson {\it et al.,} JHEP 0202(2002)037;\hfil\break
  Paul Langacker, {\tt hep-ph/0211065;}
T. Takeuchi,
 {\tt hep-ph/0209109};\hfil\break T. Takeuchi {\it et al.,} 2002
 Nagoya Workshop, 
{\tt www.eken.phys.nagoya-u.ac.jp/;}\hfil\break
 W. Loinaz, N. Okamura,
T. Takeuchi,
{\it et al.,} {\tt hep-ph/0210193, hep-ph/0304004}.

\noindent[11] S.L. Glashow, {\tt hep-ph/0301250};\hfil\break
W. Loinaz {\it et al.,} {\tt hep-ph/0304044.}
\bye